\documentclass[reprint,amsmath,amssymb,aps,floatfix,
prl,
]{revtex4-2}

\usepackage[utf8]{inputenc} % Umlaute direkt verwenden, load before csquotes
\usepackage[babel]{csquotes} % context sensitive quotation, recommended with biblatex

\usepackage{graphicx}% Include figure files
\usepackage{dcolumn}% Align table columns on decimal point
\usepackage{bm}% bold math
\usepackage[]{units}	% example: \unit[val]{dim}
\usepackage{xcolor}
\usepackage{soul}
\usepackage{changes}
\usepackage{hyperref}% add hypertext capabilities
\usepackage[all]{hypcap}
\usepackage{textcomp}
\usepackage{upgreek}
\usepackage{enumerate}
\usepackage{multirow} % columns in tables can span multiple rows

\newcommand{\bra}[1]{\ensuremath{\left\langle#1\right|}}
\newcommand{\ket}[1]{\ensuremath{\left|#1\right\rangle}}

\newcommand{\abs}[1]{\ensuremath{{\left\lvert {#1} \right\rvert}}}

\begin{document}

\title{Coherence of symmetry-protected rotational qubits in cold polyatomic molecules}

\author{Maximilian Löw}
\email{maximilian.loew@mpq.mpg.de}
\author{Martin Ibrügger}
\author{Gerhard Rempe}
\author{Martin Zeppenfeld}

\affiliation{Max-Planck-Institut f\"ur Quantenoptik, Hans-Kopfermann-Strasse 1, 85748 Garching, Germany}

\date{\today}% It is always \today, today, but any date may be explicitly specified

\begin{abstract} % max. 600 characters including spaces for PRL
\noindent Polar polyatomic molecules provide an ideal but largely unexplored platform to encode qubits in rotational states. Here, we trap cold (100-600\,mK) formaldehyde (H$_2$CO) inside an electric box and perform a Ramsey-type experiment to observe long-lived ($\sim100$\,\textmu s) coherences between symmetry-protected molecular states with opposite rotation but identical orientation, representing a quasi-hidden molecular degree of freedom. As a result, the observed qubit is insensitive to the magnitude of an external electric field, and depends only weakly on magnetic fields.
Our findings provide a basis for future quantum and precision experiments with trapped cold molecules.
\end{abstract}

%Discipline: Atomic, Molecular & Optical
%Keywords: Cold and ultracold molecules, molecule trapping & guiding, rotational states

\maketitle

%Introduction
\noindent Quantum information science~\cite{Nielsen2010} has experienced a spectacular evolution over the last decades, employing both natural and artificial atoms as material qubits. However, many challenges remain, one of the biggest being the identification of systems that can process and store qubits over long times in a noisy real-world environment~\cite{Preskill2018}. Against this backdrop, polar molecules have been proposed as favorable systems early on~\cite{DeMille2002, Yelin2006, Rabl2006, Carr2009, Wei2011}, long before the techniques necessary to control them were developed. The suitability of these molecules arises from the long radiative lifetimes of rotational states in combination with their strong and long-range dipole-dipole interaction which allows for robust quantum-logic gates~\cite{Ni2018,Yu2019}. With this in mind, new techniques for the creation of trapped cold molecules have been developed over the last years~\cite{Prehn2016, Anderegg2018, Caldwell2019, Ding2020, Langin2021, Schindewolf2022}, allowing for new research in fields like quantum simulation~\cite{Syzranov2016, Bohn2017, Yao2018, Blackmore2019}, quantum information~\cite{Park2017, Albert2020, Hughes2020, Sawant2020, Gregory2021, Tscherbul2023}, precision measurements~\cite{ACME2018, Prehn2021, Grasdijk2021} or collision studies~\cite{Segev2019, Bause2021, Koller2022, Tang2023}, see also the review~\cite{Softley2023}. Breakthrough experiments with individual molecules in optical tweezers have now promoted molecules as advantageous candidates for a new quantum-computing architecture~\cite{Anderegg2019, Cheuk2020, Kaufman2021, Holland2023, Ruttley2023, Park2023, Bao2023}.

A recent proposal discusses the possibility to store and process quantum information in so-called quasi-hidden molecular degrees of freedom~\cite{Zeppenfeld23}. The paper identifies groups of degenerate molecular states which to a strong degree interact identically with the environment. As a result, the degree of freedom associated with these states can be factored out from the remaining molecular degrees of freedom, allowing it to be considered as a separate isolated system. Quantum information stored here will thereby be strongly protected from decoherence. Moreover, quantum operations can be performed independently on the remaining molecular degrees of freedom, without affecting quantum information stored in the quasi-hidden degree of freedom. As a result, quasi-hidden degrees of freedom represent a powerful tool for processing quantum information with molecules.

\begin{figure} [b]
	\centering
	    \includegraphics{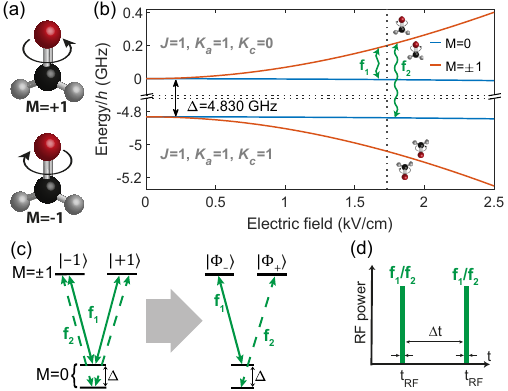}
	\caption{
		(a): Illustration of formaldehyde rotating in opposite directions. The two corresponding rotational states are perfectly degenerate in arbitrary electric fields. (b): Stark shift of the $\ket{\mathrm{J,K_a,K_c,M}}=\ket{1,1,0/1,\mathrm{M}}$ states. $\mathrm{K_a}$ and $\mathrm{K_c}$ are projections of J on the molecular axes in the limiting cases of prolate and oblate symmetric top. The vertical dotted line indicates the applied electric field for measurements with these states in this letter. Radio-frequency (RF) radiation with a frequency of either $\mathrm{f_1}$ or $\mathrm{f_2}$ couples the degenerate pair of low-field-seeking M=$\pm$1 states ($\ket{{+}1}$ and $\ket{{-}1}$) to one of the two untrapped M=0 states. (c): Emergence of bright and dark states. The RF coupling leads to the formation of a V-type three level system with a bright and dark superposition state. Molecules are transferred from the trapped bright state to the untrapped M=0 state. Bright and dark state reverse their roles when coupled to the other M=0 state. (d): RF double pulses. A single saturated RF pulse of length t$_\mathrm{RF}$ and with a frequency of either f$_1$ or f$_2$ leads to a reduced molecular signal. A second pulse after a varying time $\Delta\mathrm{t}$ probes for decoherence effects.
	}
	\label{fig:fields}
\end{figure}

A particularly appealing possibility for realizing a quasi-hidden degree of freedom is to consider the symmetry protected qubit formed by opposite angular momentum partner states under time reversal symmetry, i.e. state pairs with effective spin quantum number $\pm \mathrm{M}$. Such pairs correspond to the molecule either rotating clockwise or counterclockwise, see Fig.~\ref{fig:fields}(a), while its orientation in space is the same. The symmetry-protected nature of such qubits can be formalized by noting that both the time reversal operator $\hat{T}$ and a given component of the total angular momentum $\hat{J}_z$ commute with the molecule Hamiltonian $\hat{H}_{mol}$, whereas $\hat{T}$ and $\hat{J_z}$ anticommute. As a result, for any simultaneous eigenstate $\ket{\Psi}$ of $\hat{H}_{mol}$ and $\hat{J}_z$ with nonzero eigenvalue for $\hat{J}_z$, the states $\ket{\Psi}$ and $\hat{T}\ket{\Psi}$ form a qubit with in principle infinite coherence under the molecule Hamiltonian.

The coherence of opposite-rotation state pairs is preserved for any external interaction Hamiltonian $\hat{H}_{int}$ which also commutes with both $\hat{T}$ and $\hat{J_z}$. In particular, this is the case for interactions with an arbitrary time varying electric field along the z-direction. As a result, opposite-rotation state pairs are extremely insensitive to electric field fluctuations. More generally, opposite-rotation state pairs are insensitive to any interaction with the molecular electric dipole moment along the z-direction. This would allow the remaining molecular degrees of freedom to be entangled with an external quantum system via, e.g., dipole-dipole interactions, without affecting a qubit stored in the opposite-rotation state pairs. In contrast, magnetic-field interactions maximally break the symmetry of our symmetry-protected states. In molecules with unpaired electrons, this severely compromises the usefulness of opposite-rotation state pairs as a protected qubit. However, for closed-shell molecules, magnetic field interactions are suppressed by several orders of magnitude, allowing long coherence times to be achieved with modest control of ambient magnetic fields.

Here, we experimentally demonstrate opposite-rotation state pairs as symmetry-protected molecular qubits. We prepare superposition states for such qubits in cold formaldehyde molecules stored inside an electric trap, and observe coherence times up to 100\,\textmu s, limited by molecules moving out of resonance with the radiation fields needed for preparation and detection of the coherence. By applying a magnetic field of about $15\,$Gauss inside the electric trap, we observe quantum-beat oscillations between even and odd superposition states. As discussed below, our electric trap represents a non-ideal environment to observe very long coherence times between molecular states. However, the mere ability to nonetheless observe coherence demonstrates the robust nature of our qubits against electric-field variations.

We use a microstructured electric trap for our experiment, providing a homogeneous electric offset field in a substantial fraction of the trap volume. The trap and most of the surrounding experimental setup have been described in previous publications~\cite{Englert2011,Glockner2015,Prehn2016,Prehn2021}. A novelty compared to previous work is the molecule-detection setup based on laser-induced fluorescence (LIF), allowing the selective detection of different rotational states of formaldehyde, even down to individual M-sublevels~\cite{Ibrugger2021}. A detailed description of our setup including the LIF detection is provided in the Supplemental Material~\cite{Supplement}.

A main challenge in our experiment is the preparation and the detection of superpositions of suitable symmetry-protected qubit states.
The lowest-energy symmetry-protected states in formaldehyde which can be stored in our trap are the states $\ket{\mathrm{J,K_a,K_c,M}}=\ket{\mathrm{1,1,0,\pm1}}$, where J is the total rotational angular momentum, $\mathrm{K_a}$ and $\mathrm{K_c}$ are projections of J on the molecular axes of inertia in the limiting cases of prolate and oblate symmetric top, and M is the projection of J on the direction of the electric field. The Stark shifts of these states are depicted in Fig.~\ref{fig:fields}(b). The limited optical access to our electric trap severely restricts the opportunities to apply electromagnetic radiation with a well-defined polarization, as required to prepare and detect specific superpositions. 

A solution to this problem is offered by the two close lying M=0 states, which are energetically distinct due to the slightly asymmetric rotational structure of formaldehyde, which leads to K-type doubling. When one of these two states is coupled to the degenerate low-field-seeking M=$\pm$1 states -- from now on called $\ket{{+}1}$ and $\ket{{-}1}$ -- by a resonant radio-frequency (RF) field, this effectively forms a V-type three-level-system, where one ground state is coupled to two separate excited states as illustrated in Fig.~\ref{fig:fields} (c). In such a situation there is always a linear superposition of the two excited states unaffected by the RF, the so-called dark state, while the orthogonal superposition -- the bright state -- is coupled to the ground state and a transition between the two can be driven~\cite{Scully1997}. The exact composition of bright and dark state, which we call $\ket{\Phi_{\pm}}$, depends on the polarization of the RF radiation and is described in the Supplemental Material~\cite{Supplement}. In the case of linear polarization, the $\ket{{+}1}$ and $\ket{{-}1}$ states contribute equally and form either an even or an odd superposition state. The bright and dark states reverse their roles when coupled to the other M=0 state and can thereby be selectively addressed depending on which one of the two transition frequencies $\mathrm{f_1}$ and $\mathrm{f_2}$ is chosen.

To probe for coherence, we prepare molecules in the states $\ket{{+}1}$ and $\ket{{-}1}$ in our electrostatic trap: A single broadband saturated RF pulse of length t$_\mathrm{RF}$ roughly equilibrates the population between the addressed $M=0$ state and molecules originally in the bright superposition state, while the population in the dark state remains unchanged. As molecules transferred to M=0 are no longer trapped, they get lost from the trap once they reach its boundaries after a few hundred microseconds, which leads to a decrease in the measured signal. We refer to this decrease as molecular depletion in the following. After an adjustable time $\Delta\mathrm{t}$ we add a second RF pulse to probe for decoherence effects, as shown in Fig.~\ref{fig:fields} (d). After the application of such pulse pairs we measure the combined remaining population in both qubit states. Since the first pulse already equilibrates the population between the $M=0$ state and the bright superposition state, we expect the second pulse to cause substantially less molecule depletion for short $\Delta\mathrm{t}$. However, decoherence between the bright and the dark superposition state during the time $\Delta\mathrm{t}$ will allow the second pulse to transfer additional molecules to the $M=0$ state, resulting in additional depletion.

\begin{figure}[b]
	\centering
	\includegraphics{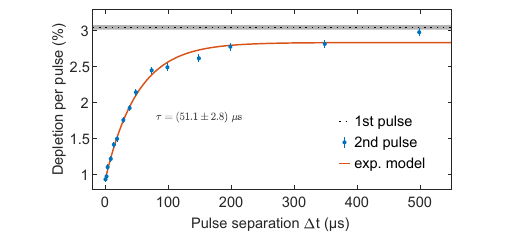}
	\caption{
		Radio-frequency depletion of molecules. H$_\mathrm{2}$CO molecules in the states $\ket{1,1,0,\pm1}$ and at a temperature of 600\,mK are coupled by pairs of 5\,\textmu s long RF pulses to the untrapped state $\ket{1,1,0,0}$. The amount of depletion caused by the second pulse of each pair is clearly dependent on the pulse separation time $\Delta\mathrm{t}$ and approaches the value of the first pulse (indicated by the gray dashed line) on a time scale $\tau$ of $51.1\pm2.8$\,\textmu s.
	}
	\label{fig:dpulse_nob}
\end{figure}

The results for a double pulse experiment as described are shown in Fig.~\ref{fig:dpulse_nob}.
Here, a sample of molecules in the states $\ket{1,1,0,\pm1}$ (i.e. $\ket{{+}1}/\ket{{-}1}$) stored in the trap at a temperature of about 600\,mK interacts with pairs of RF pulses of frequency $\mathrm{f_1}$ separated by a varying time $\Delta\mathrm{t}$. Afterwards, the molecules are unloaded over several seconds into the LIF detection setup, where the measured total signal consists of the fluorescence signal of molecules in $\ket{{+}1}/\ket{{-}1}$ integrated over the unloading time. Details of the sample preparation and the experimental parameters are provided in the Supplemental Material~\cite{Supplement}.

We observe a clear dependence of the depletion caused by the second pulse on $\Delta\mathrm{t}$. For short time separations, the second pulse causes substantially less depletion than the first pulse. For larger $\Delta \mathrm{t}$ this increases until the depletion of both pulses is roughly the same. This increase can be very well described by an exponential model with an extracted decay constant $\tau$ of about 50\,\textmu s. For time separations significantly longer than $\tau$ the two pulses are effectively independent of each other.

The reduced depletion for short $\Delta\mathrm{t}$ indicates that both RF pulses are interacting with the same molecules in the bright superposition state, whereas molecules in the dark superposition state are expected to be left unaffected. However, the increased depletion for larger $\Delta\mathrm{t}$ is not primarily due to decoherence between the bright and the dark state, but instead due to movement of the molecules in the trap. Since the electric field in the trap increases roughly exponentially towards the edge of the trap, at any moment in time only a fraction of the stored molecules is located in the trap center region where it is on resonance with the RF pulse. This can also be seen from the small amount of depletion ($\sim3\%$) caused by a single saturated pulse. As the molecules can move around freely in the trap, those affected by a given pulse move out of the trap center and out of resonance while simultaneously other molecules, that did not interact with this pulse, come into it. Consequently, for sufficient time separation $\Delta\mathrm{t}$ between the two pulses, the second pulse interacts with an essentially completely new sample of molecules.
Therefore, the extracted decay constant can be seen as a lower limit on the coherence time between the bright and dark states. However, we would expect a similar result for a very fast decoherence process between the bright and dark states (with a decoherence time shorter than the pulse length) or if the dark state didn't exist at all, although these two possibilities can basically be excluded based on theory.

To rule out any alternative interpretations, we apply a magnetic field to the molecules inside the trap. The field is created by a coil placed on top of our vacuum chamber and consequently oriented perpendicularly to the capacitor plates of the electric trap, roughly along the offset field $\mathcal{E}_{\mathrm{offset}}$. The field lifts the degeneracy of the $\ket{{+}1}/\ket{{-}1}$ states due to Zeeman splitting, which is known to be on the order of a few kHz/G for formaldehyde~\cite{Huttner1968}.  Consequently, coherent superpositions of $\ket{{+}1}/\ket{{-}1}$ become time-dependent as Larmor precession occurs. 
In our measurement scheme, the first RF pulse creates a population imbalance between the bright and the dark superposition state, with a population surplus in the latter. With the external magnetic field applied, the population surplus oscillates between the two states, which is probed by the second pulse.

\begin{figure}[tb]
	\centering
	\includegraphics{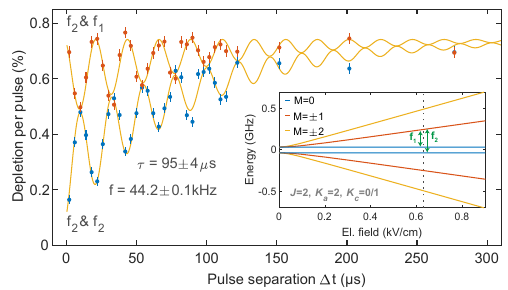}
	\caption{
		Coherence between symmetry-protected states. Molecules at 100\,mK are prepared in states $\ket{2,2,0,\pm1}$. A first pulse of length 2\,\textmu s and $\mathrm{f_2}$ couples one of the two superposition states of $\ket{2,2,0,\pm1}$ with the $\ket{2,2,1,0}$ state. After a time interval $\Delta \mathrm{t}$, a second pulse of the same length is applied, either with the same frequency (blue data points, bottom) or a frequency of $\mathrm{f_1}$ (red data points, top). The measurement is performed with a magnetic field applied to the molecules causing precession between the superposition states of $\ket{2,2,0,\pm1}$ resulting in oscillations in the depletion signal caused by the second pulse. A time constant is extracted from a fitting function to the oscillations. The inset shows the Stark shift of the $\ket{2,2,0/1,\mathrm{M}}$ manifold and the driven transitions with the dashed line indicating the electric field applied in this measurement.
	}
	\label{fig:dpulse_b}
\end{figure}

In addition to adding a magnetic field, we switch to different states acting as $\ket{{+}1}/\ket{{-}1}$, namely $\ket{\mathrm{J,K_a,K_c,M}}=\ket{2,2,0,\pm1}$. This allows us to showcase a different pair of symmetry protected qubit states and, more important, to address the superpositions of $\ket{{+}1}$ and $\ket{{-}1}$ with RF pulses of frequency $f_2$. We note that for the $\ket{1,1,0,\pm1}$ states used so far, the large inversion splitting between the $\mathrm{K_c}$=0 and $\mathrm{K_c}$=1 states of $4.83$\,GHz at zero field prevents us from applying saturated microwave pulses of frequency $\mathrm{f_2}$ with sufficiently short duration due to technical limitations. For the J=$\mathrm{K_a}$=2 states, however, the splitting only amounts to approximately 71\,MHz. This is sufficiently large that it allows the transitions from the M=$\pm1$ states to the two M=0 states as well as to the M=$\pm2$ states to be resolved and driven independently, but small enough that both the $\mathrm{f_1}$ and the $\mathrm{f_2}$ transition can now be saturated with short RF pulses. The Stark shift for these states is shown in the inset of Fig.~\ref{fig:dpulse_b}.

Furthermore, to reduce the motion-induced effects described above, we create a colder sample of molecules with a temperature of about 100\,mK by a combination of optoelectrical Sisyphus cooling~\cite{Prehn2016} and optical pumping, see Supplemental Material for experimental details~\cite{Supplement}. For the measurement shown in Fig.~\ref{fig:dpulse_b}, the external magnetic field used is on the order of 15\,Gauss. First, a short RF pulse with frequency $\mathrm{f_2}$ is applied followed after a varying time $\Delta \mathrm{t}$ by an identical second pulse in case of the blue data points or a pulse with frequency $\mathrm{f_1}$ in case of the red data points. In both cases, damped oscillation patterns with the same frequency are clearly visible in Fig.~\ref{fig:dpulse_b}. The pattern is overlaid with the already familiar exponential decay in the case of the blue data points. As the molecules Larmor precess between the dark and bright superposition states between the two pulses, the number of molecules addressable by the second pulse varies with time, leading to oscillations in the depletion signal. The second pulse addresses either the same or the orthogonal superposition state as the first one, leading to a $\pi$-phase shift between both oscillations~\cite{Supplement}. Time scales of the exponential decay and the damping of the oscillation are statistically not significantly different, so both are described by a single parameter in the fitted model. With $95\pm4$ \textmu s the extracted decay constant is roughly twice as large as in Fig.~\ref{fig:dpulse_nob}.

The appearance of the oscillation patterns proves the presence of coherence between the bright and dark states and therefore also between the two degenerate energy states $\ket{{+}1}/\ket{{-}1}$. It is limited by the observed exponential decay, which can again be explained by the movement of the molecules. The observation of an increased decay time for this colder sample compared to the one shown in Fig.~\ref{fig:dpulse_nob} supports this interpretation. The extracted decay constant is consequently not a coherence time as such, but the timescale during which we can interact with the same molecules in the trap. 

There are, however, also sources of decoherence in our setup: As the magnetic field is created by a single coil, it is not perfectly homogeneous over the trap volume. Especially in a direction parallel to the capacitor plates, which are several centimeters long, the field varies by up to $20\%$ putting a serious limit on the possible coherence time on the order of what we observe for the stronger (several tens of Gauss) magnetic fields we apply. In addition, the experimental chamber was not assembled with a focus on magnetic properties, so the actual magnetic fields at the trap site are unknown and could add an additional source of inhomogeneity. Another decoherence mechanism arises from the spatially varying direction of the electric field near the edge of the trap. As the field direction changes along the molecules trajectory, the $\ket{{+}1}/\ket{{-}1}$ states adiabatically follow the direction of the applied electric field. However, a relative geometric phase between these states is accumulated in the process~\cite{Berry1984}. This phase depends on the molecule trajectory, and thus leads to decoherence. Effectively, this limits the coherence times to the time it takes a molecule to move from the center of the trap to the trap surface. These limitations arise from our experimental setting and can be circumvented in an environment with more control over the molecular movement and the external electric and magnetic fields.

In conclusion, we have exploited the symmetry of our polyatomic molecules to observe a coherence between rotational (clock-like) qubit states in an electric trap. The qubit is by construction insensitive to the magnitude of an applied electric field, a valuable feature for quantum-information processing, and tolerates adiabatic changes in the electric field direction if the resulting geometric phase can be taken into account. Our observation method resembles a quantum beat spectroscopy~\cite{Scully1997} that provides not only information about the response of the qubit states to an external magnetic field but that could also be used to investigate the internal molecular structure including the hyperfine interaction (which is too small to be observable in our present work).

More possible applications lie in fundamental physics. Basically the same state pairs as we have investigated here are of great value for, e.g., searches for an electric dipole moment of the electron (eEDM)~\cite{Hutzler2020, Augenbraun2021}, or nuclear Schiff moments~\cite{Yu2021}. Note, however, that distinct molecule species should be used for eEDM measurements and quantum information applications, since the former require molecules with unpaired electrons, whereas closed-shell molecules feature several orders of magnitude reduced magnetic field sensitivity for qubit storage. Nonetheless, coherence between opposite rotation state pairs has been observed for CaOH, a molecule with an unpaired electron~\cite{Anderegg2023}. Here, the magnetic field sensitivity is reduced by about two orders of magnitude by tuning it near a zero crossing via an electric offset field. Similar opportunities to further reduce the magnetic field sensitivity almost certainly exist for closed-shell molecules as well. All this opens the door to fascinating possibilities in the years to come.

\begin{acknowledgments}
The authors thank Albert Gasull for helpful discussions. This work was supported by Deutsche Forschungsgemeinschaft under Germany’s excellence strategy via Munich Center for Quantum Science and Technology EXC-2111-390814868. M.Z. acknowledges support from the Deutsche Forschungsgemeinschaft via Grant No. ZE 1096/2-1.
\end{acknowledgments}
M.L. and M.I. contributed equally to this work.

\clearpage

% Supplementary Material
% change figure labels to FIG. S... and reset counter
\mathchardef\mhyphen="2D
\renewcommand{\thefigure}{S\arabic{figure}}
\renewcommand{\theHfigure}{S\arabic{figure}} % for hyperref
\setcounter{figure}{0}

\appendix
\section*{Supplemental Material}
\subsection{Composition of bright and dark state}

\noindent In a three-level system, where one ground state is coupled to two excited states via external electromagnetic radiation, there is always a superposition of the two excited states decoupled from the transition -- the so-called dark state -- while the orthogonal superposition -- the bright state -- remains coupled. In our measurement scheme this situation arises when we couple the degenerate $\ket{+1}$ and $\ket{-1}$ states to one of the two $M=0$ states with radio frequency (RF) radiation.

Formaldehyde is an asymmetric-top rotor and its rotational states can be written as a combination of symmetric rotor states \cite{Townes1975}:
\begin{equation*}
    \ket{\Phi_{J,K_a,K_c,M}}=\sum_{K=K_a+2n}{a_{J,K,M}\ket{\Psi_{J,K,M}}}
\end{equation*}
with $n\in\mathbb{Z}$ s.t. $\abs{K}\leq J$.
As formaldehyde is only slightly asymmetric, the main contributions come from the $K=\pm K_a$ states and we can in good approximation neglect the other terms. In the following we limit ourselves to cases with $K_a=J$ and $K_c=0/1$ for easier readability, but our considerations are equally valid for other states as well. The degenerate states $\ket{+1}=\ket{\Phi_{J,J,0,1}}$ and $\ket{-1}=\ket{\Phi_{J,J,0,\mhyphen1}}$ used in this work can be written as:
\begin{equation*}
    \ket{\Phi_{J,J,0,1}}=\frac{1}{\sqrt{2(1+\epsilon^2)}}[(1-\epsilon)\ket{\Psi_{J,J,1}}-(1+\epsilon)\ket{\Psi_{J,\mhyphen J,1}}]
\end{equation*}
\begin{equation*}
    \ket{\Phi_{J,J,0,\mhyphen1}}=\frac{1}{\sqrt{2(1+\epsilon^2)}}[(1+\epsilon)\ket{\Psi_{J,J,\mhyphen1}}-(1-\epsilon)\ket{\Psi_{J,\mhyphen J,\mhyphen1}}]
\end{equation*}
with a parameter $0\leq\epsilon\leq1$ dependent on a static external electric field ($\epsilon=0$ for zero field). The two $M=0$ states involved are:
\begin{equation*}
    \ket{\Phi_{J,J,0,0}}=\frac{1}{\sqrt{2}}[\ket{\Psi_{J,J,0}}-\ket{\Psi_{J,-J,0}}]
\end{equation*}
\begin{equation*}
    \ket{\Phi_{J,J,1,0}}=\frac{1}{\sqrt{2}}[\ket{\Psi_{J,J,0}}+\ket{\Psi_{J,-J,0}}]
\end{equation*}
When the degenerate $\ket{+1}$ and $\ket{-1}$ states are coupled to one of the $M=0$ states, the transition matrix elements can be derived from those of a symmetric top, which in turn are determined via the direction cosine matrix elements~\cite{Townes1975}. The elements relevant here are equal to:
\begin{equation*}
   \bra{\Psi_{J,K,0}}\hat{e}_x\vec{d}\ket{\Psi_{J,K,\pm1}}=\frac{d K}{2\sqrt{J(J+1)}}
\end{equation*}

\begin{equation*}
   \bra{\Psi_{J,K,0}}\hat{e}_y\vec{d}\ket{\Psi_{J,K,\pm1}}=\pm i\frac{d K}{2\sqrt{J(J+1)}}
\end{equation*}
with $\vec{d}$ being the electric dipole operator and $d$ the magnitude of the molecular electric dipole moment.
We assume the RF field is linearly polarized in an arbitrary direction not parallel to the quantization direction along the $z$-axis defined by the trap offset field $\mathcal{E}_{\mathrm{offset}}$. We only consider the components perpendicular to the field, as only those are relevant for transitions with $\Delta M=\pm1$:
\begin{equation*}
    \vec{E}_{RF}=E_{RF}\cos{\varphi}\;\hat{e}_x+E_{RF}\sin{\varphi}\;\hat{e}_y
\end{equation*}\\
with $0\leq\varphi\leq2\pi$. The relevant transition matrix elements then amount to:
\begin{equation*}
    \bra{\Phi_{J,J,0,0}}\vec{d}\cdot\vec{E}_{RF}\ket{\Phi_{J,J,0,\pm1}}=\mp A \epsilon e^{\pm i \varphi}
\end{equation*}
\begin{equation*}
    \bra{\Phi_{J,J,1,0}}\vec{d}\cdot\vec{E}_{RF}\ket{\Phi_{J,J,0,\pm1}}= A e^{\pm i \varphi}
\end{equation*}
with $A$ being a constant dependent on the values of $J$, $\mathcal{E}_{\mathrm{offset}}$, $E_{RF}$ and $d$. Now it is straightforward to see that the two orthogonal superposition states
\begin{equation*}
    \ket{\Phi_{\pm}}=\frac{1}{\sqrt{2}}[\ket{\Phi_{J,J,0,1}}\pm e^{i2\varphi}\ket{\Phi_{J,J,0,\mhyphen1}}]
\end{equation*}
are a bright and a dark state when coupled to $\ket{\Phi_{J,J,1,0}}$ and vice versa when coupled to $\ket{\Phi_{J,J,0,0}}$.

\subsection{Experimental setup}

\begin{figure}[t]
	\centering
	\includegraphics{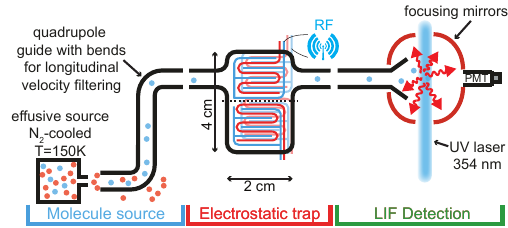}
	\caption{
		Experimental setup. Formaldehyde molecules from a thermal source (T $\approx$ 150\,K) are velocity filtered and loaded into the trap via an electric quadrupole guide. In the electrostatic trap RF radiation can be applied to the molecules. Via a second guide they are transferred to the detection area, where a UV laser drives a vibronic transition and the emitted photons are focused onto a photomultiplier tube (PMT).
	}
	\label{fig:sequence}
\end{figure}

\noindent Our experimental setup is depicted schematically in Fig.~\ref{fig:sequence}. We store velocity-filtered formaldehyde molecules in an electrostatic trap that -- as described previously~\cite{Englert2011} -- consists of a pair of microstructured capacitor plates and a perimeter electrode. The resulting multipole electric field configuration has strong electric fields near the electrodes (nominal trap depth of $\sim$1K) and a tunable fairly homogeneous offset electric field $\mathcal{E}_{\mathrm{offset}}$ in the trap center. Effectively, this creates a three-dimensional box-like potential for molecules in low-field-seeking states. The quantization axis of our system is defined by the direction of $\mathcal{E}_{\mathrm{offset}}$. Its main component is caused by an average voltage difference between the two capacitor plates and therefore oriented perpendicularly to them. However, the individual stripes of the microstructure electrodes are wedge-shaped to reduce the amount of electric-field zeros in the trap, which would otherwise severely limit the trapping time. This results in an additional field component parallel to the plates. A linearly polarized electromagnetic field between the two capacitor plates can be created by applying RF radiation directly to them leading to a purely perpendicular polarization direction with respect to the plates. Therefore, the polarization direction and the offset field $\mathcal{E}_{\mathrm{offset}}$ are not aligned allowing the driving of $\Delta M=\pm1$ transitions as there is a polarization component perpendicular to the quantization axis.

The storage time in the trap amounts to several seconds, depending on the molecular state and kinetic energy, and can be increased to about a minute by cooling the molecules via optoelectrical Sisyphus cooling~\cite{Prehn2016}. This gives plenty of time to perform experiments after which the molecules are unloaded from the trap via an electric quadrupole guide and brought to the detection area. Here, they are observed using laser-induced fluorescence (LIF)~\cite{Ibrugger2021}: A laser with a wavelength of 354\,nm drives the $\tilde{A}^1A_2\leftarrow\tilde{X}^1A_1\ 4^1_0$ vibronic transition of formaldehyde. The photons emitted by the molecules are collected by two hollow mirrors and are focused onto a broadband photomultiplier tube (PMT). In this way, different rotational states of formaldehyde can be detected selectively by changing the excitation wavelength. When applying an electric field to the detection area, even individual M-sublevels can be resolved.

\subsection{State preparation and experimental details}

\noindent In this work, the sample of molecules in the states  $\ket{1,1,0,\pm1}$ is obtained by loading the electric trap with velocity filtered molecules for a few seconds, where some of them occupy the desired states already. This results in a sample with a temperature of about 600\,mK and a mean velocity of $(12.9\pm0.2)$\,m/s in the unloading guide. Molecules in other low-field-seeking states are also present in the trap in this case, but because of our state-selective detection they don't interfere with the measurement. The molecules are stored in the trap for a time $\mathrm{t}_\mathrm{store}$ of 50\,ms, during which the RF pulses are applied. The offset electric field during this time is chosen to be $\mathcal{E}_{\mathrm{offset}}$=$1.73$\,kV/cm, leading to a transition frequency $\mathrm{f_1}$ of 211.5\,MHz. The pulse length of the RF pulses is 5\,\textmu s.
To increase the overall depletion signal, we apply multiple pairs of pulses separated by a waiting time $\mathrm{t}_\mathrm{wait}$, which is long enough to allow that two pulse pairs don't interfere with each other. In total, N=20 pulse pairs are applied during $\mathrm{t}_\mathrm{store}$.

The sample of $\ket{2,2,0,\pm1}$ molecules is obtained by preparing colder molecules primarily in the states $\ket{4,4,0,4}$ and $\ket{5,4,1,5}$ using optoelectrical Sisyphus cooling and then transferring them to $\ket{2,2,0,\mathrm{M}}$ by optical pumping via a vibrational transition. Afterwards, the RF transition between $\ket{2,2,0,\pm1}$ and $\ket{2,2,0,\pm2}$ is driven for 50\,ms to increase the population in $\mathrm{M}=\pm1$ (i.e. $\ket{+1}/\ket{-1}$). In this way, we create a sample with a temperature of about 100\,mK and a mean velocity of $(5.3\pm0.3)$\,m/s in the unloading guide, which is stored in the trap for 0.5\,s. $\mathcal{E}_{\mathrm{offset}}$ is chosen to be $0.63$\,kV/cm, leading to transition frequencies of $\mathrm{f_1}$=217.5\,MHz and  $\mathrm{f_2}$=288\,MHz. The external magnetic field used is on the order of 15\,Gauss and the RF pulse length 2\,\textmu s. Similar as before, N=50 independent pairs of pulses interact with the molecules to increase the overall depletion signal. In both cases the RF pulses are broadened to roughly 10 MHz by frequency modulation to increase the amount of addressed molecules per pulse.

\newpage

\end{document}